\documentstyle[epsfig,a4,11pt]{article}
\begin{document}
\def \lta {\mathrel{\vcenter
     {\hbox{$<$}\nointerlineskip\hbox{$\sim$}}}}
\def \gta {\mathrel{\vcenter
     {\hbox{$>$}\nointerlineskip\hbox{$\sim$}}}}

\begin{center}

{\Large \bf  Quintessential Enhancement of  Dark Matter Abundance}
\vskip1cm

{\large Francesca Rosati} 

\vskip0.3cm

{\small \it Dipartimento di Fisica `Galileo Galilei' - 
Universit\`a di Padova, via Marzolo 8, 35131 Padova - Italy} 

{\small \it INFN - Sezione di Padova, via Marzolo 8, 35131 Padova - Italy}

\vskip0.3cm

{\small E-mail:}
{\small \tt francesca.rosati@pd.infn.it}

\vskip 1cm

{\bf Abstract} 
\end{center}

The presence of a dynamical scalar field in the early universe could 
significantly affect the `freeze-out' time of particle species. 
In particular, it was recently shown that an enhancement of the relic 
abundance of neutralinos can be produced in this way.
We examine under which conditions this primordial scalar field could be 
identified with the Quintessence scalar and find, through concrete examples,
that modifications to the standard paradigm are necessary.
We discuss two possible cases: the presence of more scalars and the
switching on of an interaction.

\vskip0.8cm 
\noindent
PACS: 98.80.Cq \hspace{1cm} DFPD/03/TH/10

\section{Introduction}

As it is well known, according to the standard paradigm \cite{kolb}
a particle species goes through two main 
regimes during the cosmological evolution. 
At early times each constituent of the universe is in thermal equilibrium, 
a condition which is maintained until the particle interaction rate $\Gamma$ 
remains larger than the expansion rate $H$. 
At some time, however, $H$ will overcome $\Gamma$ because the particles will be so
diluted by the expansion that they will not interact anymore.
The epoch at which $\Gamma=H$ is called `freeze-out', and after that time 
the number of particles per comoving volume for that given species will stay
constant for the remaining cosmological history. This is how cold dark matter
particle relics (neutralinos, for example) are generated. 
This framework, although simple in principle, can be very complicated in 
practice. On one hand, we need a particle theory in order to compute $\Gamma$, 
on the other we have to choose a cosmological model to specify $H$.
Even a small change in $\Gamma$ and/or $H$ would result in a delay or 
anticipation of the `freeze-out' time of a given particle species and, 
as a consequence, in a measurable change in the relic abundance observed today.

The standard scenario \cite{kolb} assumes that before Big Bang Nucleosynthesis 
(BBN), the dominant cosmological component was radiation and so the 
Hubble parameter was evolving according to $H^2 \sim \rho_r \sim a^{-4}$, 
where $\rho_r$ is the energy density of radiation and $a$ is the scale 
factor of the universe.
This is a reasonable assumption, but the available data do not exclude  
modifications to this scenario.
For example, it is conceivable that in the pre-BBN era, a scalar field 
had dominated the expansion for some time\footnote{We mean, of course, 
after the end of inflation.}, leaving room to radiation only afterwards.
To be more concrete, if we imagine to add a significant fraction of scalar 
energy density to the background radiation at some time in the past, 
this would produce  a variation in $H^2$, depending 
on the scalar equation of state $w_\phi$.\footnote{Remember that the energy 
density of each cosmological component scales as 
$\rho_x/\rho_{x}^{o} = (a/a_o)^{-3(w_x +1)}$, if $w_x$ is the corresponding equation 
of state.} 
If $w_\phi > w_r=1/3$, the scalar energy density  
would decay more rapidly than radiation, but temporarily increase the global
expansion rate. This possibility was explicitly considered in 
Ref.~\cite{salati}, where it was calculated that a huge enhancement of 
the relic abundance of neutralinos could be produced in this way.

The effect of an early scalar field dominance on electroweak baryogenesis is discussed
in Ref.\cite{joyce}. An alternative possibility for non-standard `freeze-out' 
is proposed in Ref.\cite{max}.

In this paper we will recall how a period of scalar `kination' (see below)
could affect the relic density of neutralinos  and discuss if
this primordial scalar field could be identified with the Quintessence scalar,
{\it i.e.}~the field thought to be responsible for the present acceleration 
of the universe \cite{quint-review}.
We will find that modifications to the standard Quintessence paradigm are 
necessary and discuss some concrete examples.

\section{Scalar field `kination'}

The early evolution of a cosmological scalar field $\phi$ with a runaway
potential $V(\phi)$ is typically 
characterised by a period of so--called `kination' \cite{stein,nnr,macorra},
during which the scalar energy density
\begin{equation}
\rho_{\phi} \equiv \frac{\dot{\phi}^2}{2} + V(\phi)
\end{equation}
is dominated by the kinetic contribution $E_k= \dot{\phi}^2/2 \gg V(\phi)$.
After this initial phase, the field comes to a stop and remains nearly 
constant for some time (`freezing' phase), until it eventually reaches an 
attractor solution\footnote{For a detailed 
discussion of the existence and stability of attractor solutions for general 
potentials see \cite{nnr, macorra}.}.
A simple and interesting example is that of inverse power law potentials:
\begin{equation}
V(\phi) = M^{4+n}~\phi^{-n}
\end{equation}
with $M$ a mass scale and $n$ a positive number. These 
potentials exhibit the attractive feature of a stable attractor solution 
characterised by a constant scalar equation of state \cite{stein,attr}
\begin{equation}
w_{n} = \frac{n w -2}{n +2}
\end{equation}
which depends only on the exponent $n$ in the potential and on the 
background equation of state $w$. Since $n$ is positive, the condition
$w_{n}<w$ always holds  and the scalar field, which can be sub-dominant at
the beginning, will eventually overtake the background energy density. 
This is a welcome feature if we are modelling the present acceleration of the 
universe through the scalar field dynamics (Quintessence), since during 
matter domination ($w=w_m=0$) the attractor
has negative equation of state for any $n$.

\begin{figure}[th!]
\begin{center}
\parbox{16cm}{ \includegraphics[width=16cm]{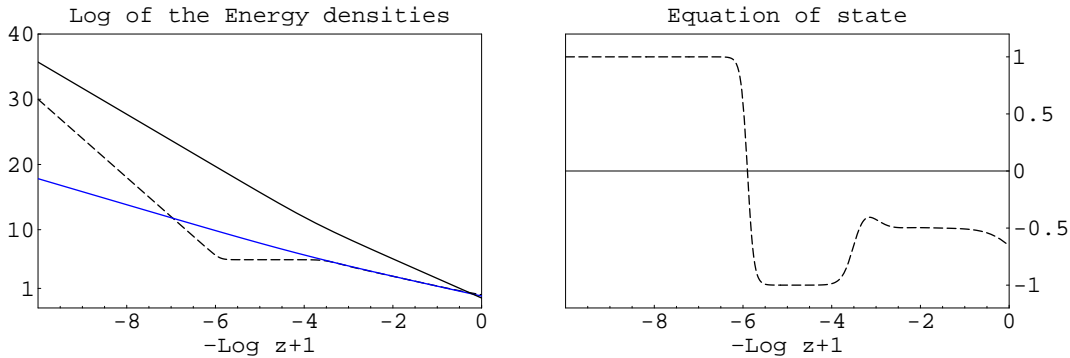}
\caption{\small The figures show the typical evolution of the relevant energy 
densities (w.r.t.~the present critical energy density $\rho_{c}^{o} \simeq
10^{-47} GeV^{4}$)
and of the scalar equation of state, for a cosmological scalar field with 
potential $V \sim \phi^{-2}$. 
The black curve corresponds to the background (radiation plus matter) and 
the blue curve to the attractor.  
The dashed lines show  the scalar energy density and equation of state: 
it can be easily seen that after an initial stage of `kination'
($w_{\phi}=1$), the field is `freezing' ($w_{\phi}=-1$) and subsequently 
joins the attractor until it overtakes the background energy density. 
On the attractor the scalar equation of state is $w_{\phi} = -1/2$.}}
\end{center}
\end{figure}

In general, a scalar field in a cosmological setting obeys the evolution 
equation
\begin{equation}
\ddot{\phi} + 3H\dot{\phi} + \frac{dV}{d\phi} = 0
\label{scaleq}
\end{equation}
and, for any given time during the cosmological evolution,
the relative importance of the scalar energy density
w.r.t.~to matter and radiation in the total energy density $\rho$
\begin{equation}
\rho \equiv \rho_{m} + \rho_{r} + \rho_{\phi}
\label{rho}
\end{equation}
depends on the initial conditions, and is constrained by the available
cosmological data on the expansion rate and large scale structure.
As it is well known, the cosmological expansion rate is governed by the 
Friedmann equation
\begin{equation}
H^2 \equiv \left( \frac{\dot{a}}{a} \right)^2 = \frac{8\pi}{3M_p^2}~ \rho
\end{equation}
where $\rho$ includes all the contributions of eq.(\ref{rho}),  and we have 
assumed a spatially flat universe. Then, if we modify the standard picture 
according to which only radiation plays a role in the post-inflationary era and 
suppose that at some time $\hat{t}$  the scalar contribution was 
small but non negligible w.r.t.~radiation, then at that time the 
expansion rate $H(\hat{t})$ should be correspondingly modified\footnote{We recall 
that the available data do not exclude that the scalar contribution could
have been important for some time in the pre-BBN era. However, and this 
is our point, future observations might be able to explore the consequences of 
this possibility.}.
Since during the kination phase the scalar
to radiation energy density ratio evolves like 
$\rho_{\phi}/\rho_r \sim a^{-3(w_\phi - w_r)} = a^{-2}$, 
the scalar contribution would rapidly fall off and leave room to 
radiation domination.

Is this of any interest to us? The answer is yes, because there is a clear
cosmological signature of this early phase: the relic density of neutralinos
\cite{salati}.
The reasoning goes as follows: since the fall off of $\rho_{\phi}$ is so rapid 
during kination, we can respect the BBN bounds and at the same time keep a 
significant scalar contribution to the total energy density just few red-shifts 
before.
For example, a scalar to radiation ratio $\rho_{\phi}/\rho_r=0.01$ at BBN 
($z\simeq 10^9$) would imply $\rho_{\phi}/\rho_r =0.1$ at 
$z \simeq 3.16 \times 10^{9}$ and $\rho_{\phi}/\rho_r =1$ at $z\simeq 10^{10}$,
if the scalar field is undergoing a kination phase.
As extensively discussed in \cite{salati}, calculations of the relic 
densities of dark 
matter (DM) particles are usually done under the assumption that the 
universe is dominated by radiation while they decouple from the primordial 
plasma and reach their final relic abundance. However, as we have seen, it is 
conceivable that the scalar 
energy density respects the BBN bounds and at the same time contributes 
significantly to the total energy density at the time DM particles
decouple.
Indeed, an increase in the expansion rate $H$ due to the additional scalar 
contribution would anticipate the decoupling of particle species and 
result in a net increase of the corresponding relic densities.
As shown in \cite{salati}, a scalar to radiation energy density ratio
$\rho_{\phi}/\rho_r \simeq 0.01$ at BBN would give an enhancement
of the neutralino codensity of roughly three orders of magnitude.

\section{Quintessence?}

As discussed in the previous section, the enhancement of the relic density 
of neutralinos requires that at some early time the scalar energy 
density was dominating the Universe. 
This fact raises a problem if we want to identify the scalar
contribution responsible for this phenomenon with the
Quintessence field \cite{quint-review} which (we suppose) 
accelerates the Universe today.  
Indeed, the scalar initial conditions are
crucial to establish the scalar energy density contribution at any time. 

In particular, the range of initial conditions which
give rise to a non-negligible Quintessence contribution at present
is huge but nonetheless does not include the case of a dominating
scalar field at the beginning.  In other words, the initial conditions
must be such that the scalar energy density is sub-dominant 
(or, at most, of the same order of magnitude of $\rho_r$) at the
beginning, if we want the Quintessence field to reach the cosmological
attractor in time to be responsible for the presently
observed acceleration of the expansion \cite{stein}.
For initial conditions $\rho_{\phi}\gta \rho_r$ we obtain the so-called 
`overshooting' behaviour: the scalar field rapidly rolls down the potential
and after the kination stage remains frozen at an energy density which 
is much smaller than the critical one. 
The larger is the ratio $\rho_{\phi}/\rho_r$ at the beginning, the smaller 
will be the ratio $\rho_{\phi}/\rho_{c}^{o}$ today.

There is also another situation in which the attractor is not reached in time.
If the initial conditions are such that 
$\rho_{\phi}\lta\rho_{c}^{o}$ (the initial scalar density is smaller than 
the present critical energy density), then the scalar field would remain
frozen throughout the whole cosmological history and join the attractor only
beyond the present time. In this case the ratio $\rho_{\phi}/\rho_{c}^{o}$ 
remains unchanged and smaller than one until today (this the so-called 
`undershooting' behaviour).

We should notice, however, that these rules strictly apply only 
to the standard case of a single
uncoupled field with an inverse power law potential $V \sim
\phi^{-n}$.  As shown in \cite{mpr} more complicated dynamics are
possible if we relax this hypothesis and consider more general
situations.  The presence of several scalars and/or of a
small coupling with the dark matter fields could modify the
dynamics in such a way that the attractor is reached in time even if
we started, for example, in the overshooting region.

\begin{figure}[th!]
\begin{center}
\parbox{16cm}{ \includegraphics[width=16cm]{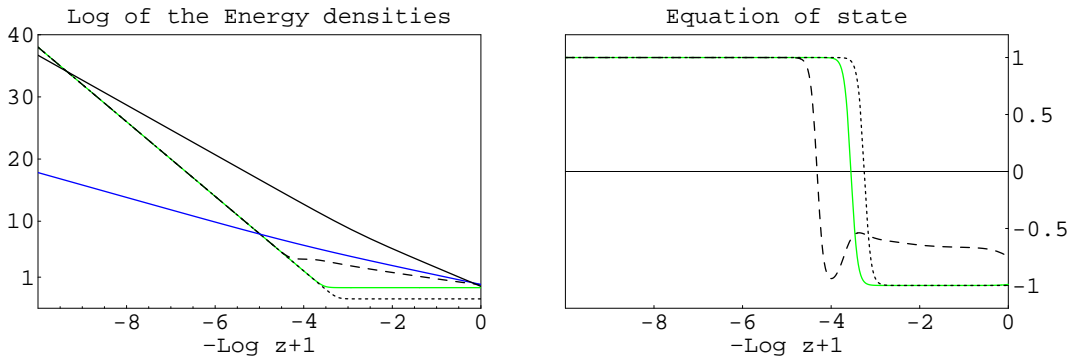}
\caption{\small The figures show the evolution of the relevant energy 
densities (w.r.t.~the present critical energy density $\rho_{c}^{o} \simeq
10^{-47} GeV^{4}$)
and of the scalar equation of state, depending on the
initial conditions, in the case of two scalar fields $\phi_{1}$ and $\phi_{2}$ 
with potential $V \sim (\phi_{1} \phi_{2})^{-1}$. The black curve
corresponds to the background (radiation plus matter) and the blue curve to the
attractor.  The green line is the case in which the two
fields start with the same initial conditions with a total energy
density corresponding to the overshooting case. If we vary the
fields' values at the beginning (keeping the total energy density
fixed), we can obtain two situations: if both the fields are still
$\ll 1$ at the beginning (dashed line) then the attractor is reached in
advance w.r.t the equal fields' case; if instead one of the fields is
$> 1$ (dotted line), then the attractor is reached later. In the
examples shown, at $z=10^{10}$ we have $\rho_{\phi}=10^{38}$ and for
the fields: $\phi_{1}=\phi_{2}=10^{-19}$ (green); $\phi_{1}=10^{-16}$, 
$\phi_{2}=10^{-22}$ (dashed) and $\phi_{1}=100$, $\phi_{2}=10^{-40}$ 
(dotted). }}
\end{center}
\end{figure}

\vskip0.5cm
\noindent
{\bf More fields.}
Consider a potential of the form
\begin{equation}
V(\phi_1,\phi_2) = M^{n+4} \left(\phi_1 \phi_2\right)^{-n/2}
\label{potential2}
\end{equation}
with M a constant of dimension mass.
In this case, as discussed first in \cite{mpr}, the two fields' dynamics 
enlarges the range of possible initial conditions
for obtaining a quintessential behaviour today.
This is due to the fact that the presence of more fields allows to play with 
the initial conditions in the fields' values, while maintaining the total 
initial scalar energy density fixed. 
Doing so, it is possible to obtain a situation in which for a fixed 
$\rho_{\phi}^{in}$ in the overshooting region, if we keep initially
$\phi_1=\phi_2$ we actually produce an overshooting behaviour, 
while if we choose to start with $\phi_1\not =\phi_2$ (and {\it the same} 
$\rho_{\phi}^{in}$) it is possible to reach
the attractor in time.
This different behaviour emerges from the fact that, if at the beginning
$\phi_2 \ll \phi_1$  then, in the example at hand,
$\partial V/\partial\phi_2 \gg \partial V/\partial\phi_1 $ and so 
$\phi_2$ (the smaller field) will run away more rapidly and tend to overshoot 
the attractor, while 
$\phi_1$ (the larger field) will move more slowly, join the attractor 
trajectory well before 
the present epoch and drive the total scalar energy density towards the 
required value\footnote{We should specify that this interplay between 
the two fields is successful only until they 
both remain smaller than 1 in Planck units.
Otherwise, also $\phi_1$ would exit the allowed region for reaching 
the attractor in time  and the behaviour of 
the total energy density would be worsened w.r.t.~the equal fields' case.
This fact can be checked numerically but has also a qualitative explanation. 
The key of the two-fields mechanism is that one field starts its cosmological evolution
far from the attractor region, while the other is kept sufficiently close in
order to join it in time. This does not work anymore if both fields start
too far away from the attractor (even though on opposite sides). 
In fact, if we allow $\phi_1$ to become larger than 1 initially, then 
this would correspond to push it down to the `undershooting' region and prevent
it as well to reach the attractor in time.}.
In figure 2 the comparison between  different cosmological evolutions 
depending on the fields' initial conditions, keeping $\rho_{\phi}^{in}$ 
fixed, is illustrated.

\begin{figure}[ht!]
\begin{center}
\parbox{16cm}{
\includegraphics[width=16cm]{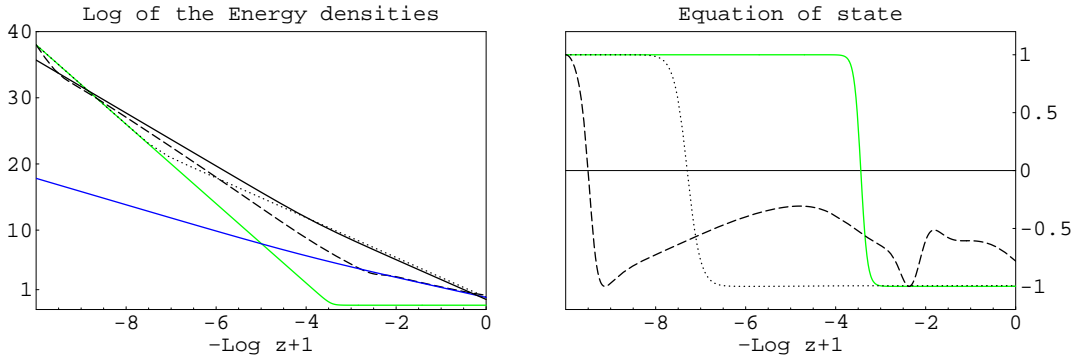}
\caption{\small\rm The figures show the evolution of the relevant 
energy densities (w.r.t.~the present critical energy density 
$\rho_{c}^{o} \simeq 10^{-47} GeV^{4}$)
and of the scalar equation of state, depending on the interaction of the
Quintessence field with the dark matter fields. The black curve corresponds to 
the background (radiation plus matter) and the blue curve to the attractor. 
The green line is the Quintessence field evolution in the 
overshooting case ($\rho_{\phi}=10^{38}$ at $z=10^{10}$), with potential
$V \sim \phi^{-2}$ and no interaction. If we switch on an additional term in
the potential, the cosmological evolution will change correspondingly.
The dashed line shows the case of a coupling 
$V_b=\frac{1}{2}bH^2\phi^2$ with 
$b=0.25$; the dotted line shows the case of a coupling 
$V_c=c\rho_m\phi$ with $c=0.5$. 
Please note that the values chosen for $b$ and $c$ in the figure are purely illustrative.
Coupling constants two orders of magnitude smaller than the ones considered here
are sufficient to ensure the desired effect. }}
\end{center}
\end{figure}

\vskip0.5cm 
\noindent
{\bf Interaction.}
Suppose now that the Quintessence scalar is not completely decoupled from the 
rest of the Universe.
Among the possible interactions, as will be discussed below, two interesting 
cases are the following:
\begin{equation}
V_b = \frac{b}{2}\ H^2 \phi^2 \;\;\;\;\;
\mbox{\rm or} \;\;\;\;\;
V_c = c \rho_m \phi
\label{inter}
\end{equation}
If we add $V_b$ or $V_c$ to the potential $V=M^{n+4}\phi^{-n}$, the 
cosmological evolution will be accordingly modified. The main effect is 
that now the potential acquires a (time-dependent) minimum and so the scalar 
field is prevented from running freely to infinity.
As a result, the long freezing phase that characterises the evolution of a 
scalar field with initial conditions in the overshooting region 
can be avoided.
As can be seen in figure 3, the interactions in eq.~(\ref{inter}) drive the 
scalar field trajectory towards the attractor (in the case of $V_b$) 
or towards $\rho_m$ (in the case of $V_c$) well before the 
non-interacting case.

Effective interaction terms like $V_b$ in eq.~(\ref{inter}) were first 
introduced in Ref.~\cite{dine} and are more recently discussed in 
Ref.~\cite{riotto}. The point is that supersymmetry 
breaking effects in the early universe can induce mass corrections to the 
scalar Lagrangian of order $H^2$. Indeed, if a term like 
$\delta K \sim \chi^* \chi \phi^* \phi$
is present in the Kahler potential, where $\chi$ is a field whose energy 
density
is dominating the universe, this will result in a correction
proportional to $\rho_\chi \phi^*\phi$ in the Lagrangian. 
Then, if the universe is critical $\rho_\chi \sim H^2$ and we obtain a 
mass correction for $\phi$ which goes like $\delta V \sim H^2 \phi^2$. 

The second type of interaction ($V_c$ in eq.~(\ref{inter})) emerges in the
context of scalar-tensor theories of gravity, in which a metric coupling
exists between matter fields and massless scalars\footnote{For a detailed
discussion of 
these theories in the context of Quintessence cosmology see, for example, 
Refs.\cite{scalar-tensor}.}. 
These theories, expressed in the so-called `Einstein frame' are defined by 
the action (see, for example, \cite{damour}):
\begin{equation}
S=S_g+S_m
\end{equation}
where $S_m=S_m[\Psi_m,A^2(\phi)g_{\mu\nu}]$ is the matter action which includes
the scalar interaction via the multiplicative factor $A^2(\phi)$ before 
the metric tensor $g_{\mu\nu}$, and the gravitational action reads
\begin{equation}
S_g = \frac{1}{16\pi G} \int d^4x \sqrt{-g} 
[R - 2g^{\mu\nu}\partial_\mu\phi\partial_\nu\phi - 2V(\phi)] \;\;\; .
\end{equation}
The scalar field equation in this context is 
modified w.r.t.~eq.(\ref{scaleq}) by the presence of an additional source term
\begin{equation}
\ddot{\phi} + 3H\dot{\phi} + \frac{1}{2}\frac{dV}{d\phi} = 
-4\pi G \alpha(\phi) T
\label{eq-bd} 
\end{equation}
where $\alpha(\phi) \equiv d\log A(\phi)/d\phi $ and $T$ is the trace of the 
matter energy-momentum tensor $T^{\mu\nu}$. The case
$\alpha(\phi)=0$ ({\it i.e.} $A(\phi)=$ const.) corresponds to the standard 
scenario with the scalar field 
decoupled from matter fields; while it can be easily seen that 
eq.(\ref{eq-bd}) is equivalent to switching on an interaction 
term like $V_c$ of eq.~(\ref{inter}), if we choose the function $A(\phi)$
to be an exponential.

As it is well known, introducing an interaction between the matter fields
and a light scalar should always done with great care in order to avoid 
unwanted effects like time variation of constants and modification of
gravitational laws 
(for discussions of these issues in the Quintessence context 
see Refs.\cite{mpr,carroll,veneziano}).
Limits on the possible values of the couplings $b$ and $c$ in eq.~(\ref{inter})
depend on the details of the theory that originates them and on the cosmological
epoch we are considering.
Just as a rough estimate, we recall that at the present time
solar system measurements impose on metric theories of gravity  
(see \cite{damour}) an upper bound for $c$ of order $10^{-1}$.

\section{Conclusions}

In this letter we have shown that modifications to the standard Quintessence 
paradigm are possible in order to make the Quintessence scalar responsible 
of an enhancement of the relic density of neutralinos. 
We have illustrated through specific examples that this can be obtained
in two different ways: by considering more scalar fields in the Quintessence 
fluid and introducing an interaction term in the scalar potential.

\vskip0.5cm
\noindent
{\bf Acknowledgements.}
It is a pleasure to thank Massimo Pietroni for useful comments and discussions.


\end{document}